# DOUBLE DIFFRACTION WHITE LIGHT IMAGING: FIRST RESULTS WITH BIDIMENSIONAL DIFFRACTION

José J Lunazzi, Noemí I Rodríguez Rivera

Laboratório de Óptica, DFMC, Instituto de Física Gleb Wataghin
Universidade Estadual de Campinas
13083-970-Campinas, SP, Brasil
E-mail: lunazzi@ifi.unicamp.br and nrivera@ifi.unicamp.br

## Abstract

The double diffraction of white light can produce a thin-prism-like image in certain conditions by using ordinary diffraction gratings. The diffractive deviation of rays happens mainly in one direction because the diffracting elements are straight and parallel. We show similar images using elements where diffraction happens in both directions: spiral grooves discs. Some differences between the straight-groove and the almost circular curved-groove cases are described.

## 1. Introduction

White-light double diffracted imaging was demonstrated in cases where a pinhole intermediated the diffraction path [1,2]. Imaging without any intermediate element was showed in a previous article [3] with second order diffraction on a first grating being collected to perform the first order diffraction by a second similar grating.  This prism-like white-light image is even more chromatically correct than a prism image. The image conditions depends on the relationship between the gratings distance and the object distance. All diffracted rays are in the same plane, it is not obvious that the same properties may arise in case of curved grooves . To test the possibility of imaging in the case of curved grooves we employed two spiral gratings within the same diffraction order relationship to try to form a similar situation. Because the curvature of the grooves made light to converge at a shorter distance than in our previous experiment we tried to find imaging with shorter distance between elements, varying the distance between diffracting elements until finding a satisfactory result [4,5]**.**

## 2. Description

A white-light object is positioned in front of a spiral diffracting element, which groove is centered to one side as in Figure 1, where the center of curvature of the groove is indicated.

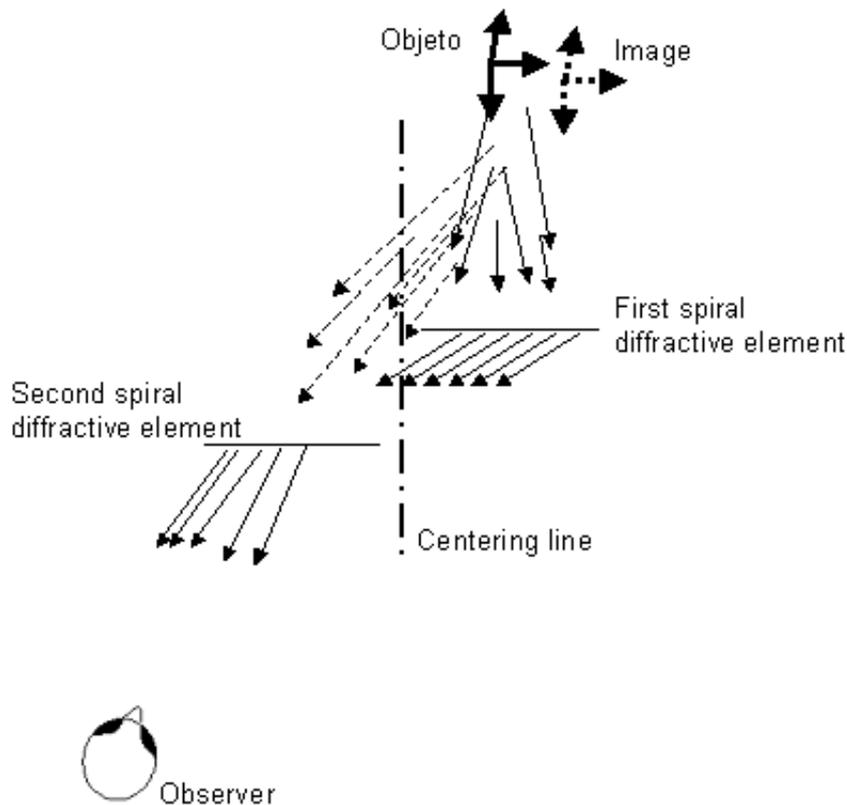

Figure 1: schematic drawing of the double diffraction imaging setup.

Only a part smaller than a quart of a complete spiral disc is employed. That is why the center of curvature is outside the diffracting element. The second diffracted order directed to the side of the center of curvature is collected at a certain distance by a second spiral diffracting element. A perpendicular to the diffracting elements through the center of one element gives a centering line for both. The deviation direction of the second diffraction is opposite to that of the first diffraction, giving a light field corresponding to a divergent image of the object.

The same ray tracing scheme of our previous article [3] can be employed to give a first primary explanation of the image formation. It is shown in Figure 2 and describes the main orientation of the diffracted light for rays that impinge perpendicularly to the tangent of the groove profile on the first diffracting element.

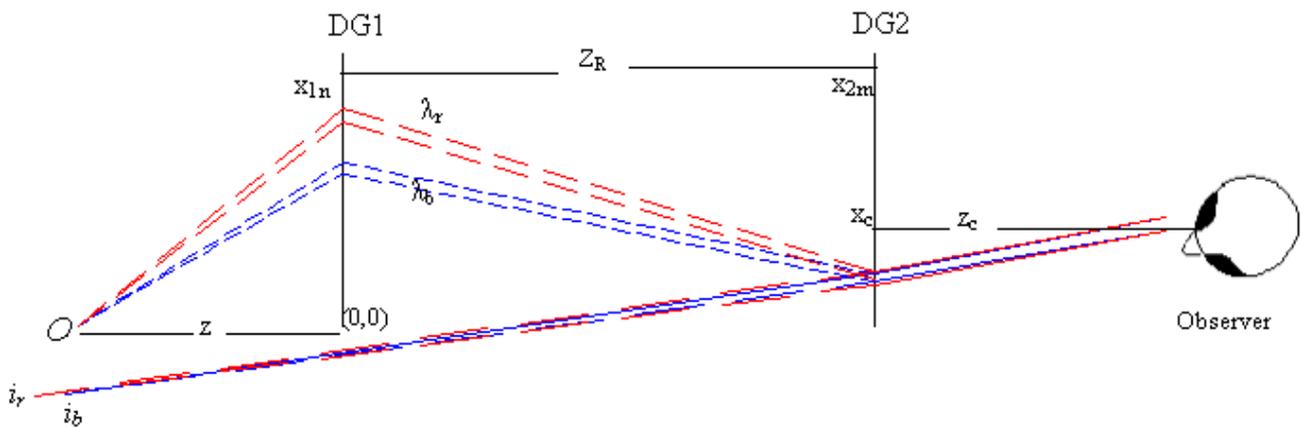

Figure 2: Ray-tracing scheme for incident perpendicular rays.

The diffraction of these rays impinge the second diffracting element in a way that keeps the plane of propagation after the second diffraction. A sequence of propagation planes can then be assumed to happen, containing the centering line and all of them performing images in the same way as ordinary gratings did [3]. Although this explanation does not include the case of rays that impinge in other directions, that is, we are excluding to consider conical diffraction, we believe that our first approach is a good way to start analyzing the phenomena.

## 3. Experimental details

A sector of less than one quarter of a compact (CD) data disc made of a single spiral groove was employed for the first diffracting element, and a similar one for the second. We name "grooves" the different appearences of this groove in a radial direction. The number of grooves per milimeter was 658. The object was a flashlight with a vertical arrow pointing upwards positioned at 15.5 cm in front of the first element  The second diffracting element was at a distance of 9.0  cm.  Figure 3 is a color photograph of the setup as seen from above, with the flashlight to the left and the two diffracting discs to the right.

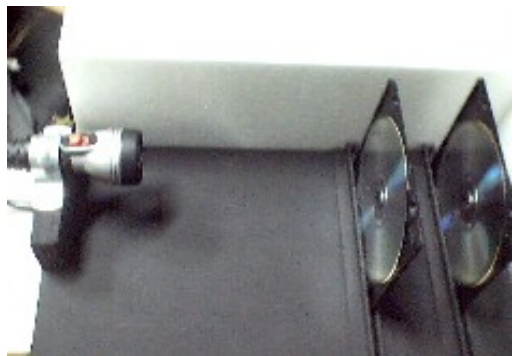

Figure 3: Color photograph of the experimental setup seen from above.

Figure 4 shows the setup from behind, in a viewpoint very close to the flashlight. It allows to see the circular holes on the first and second black holders of the diffracting discs, showing that only a small part of each disc is being employed.

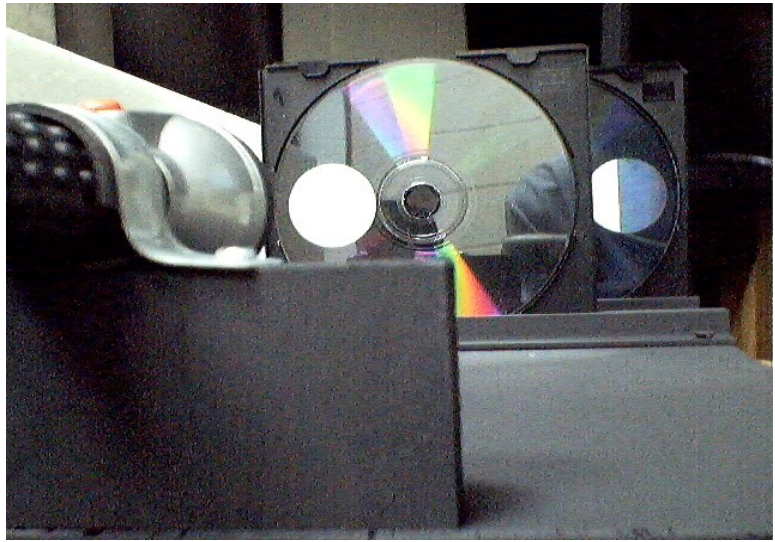

Figure 4: Color photograph of the experimental setup seen from behind.

Figure 5 shows the object in a frontal view and to its right appears the image showing the arrow vertically inverted. The same image in gray tonality was added to the right to improve the quality. Because we did not succeed to photograph at the best possible position, a color blurring appeared.

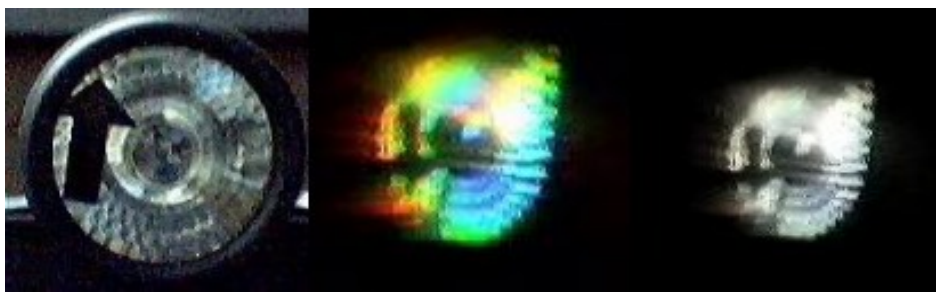

Figure 5: The object (left), its image (center), and the image in gray (right)

An interesting fact appears when comparing this experience with our previous one [1]: the depth of field for the object is almost complete, no limitations were found and it was even possible to observe scenes far away from the system.

## 4. Conclusions

We showed that it is possible to produce prism-like double diffracted white-light images when diffraction happens at different planes according to the inclination of spiral grooves. The differences with the straight-grooves case can be observed in the inversion up-down of the image. Some reduction in sharpness and a very extended longitudinal field of view was noticed.

## Acknowledgements

The "Pro-Reitoria de Pós Graduação" of Campinas State University - UNICAMP is acknowledged for a BIG fellowship for Noemí I. Rodríguez Rivera. The authors thank the Found of Assistance to Teachingand research- FAEP of Campinas State University four founding and support.